# Three-dimensional laser nanolithography of laser crystals

Airán Ródenas[1]*, Min Gu[2], Giacomo Corrielli[1], Petra Paiè[1], Sajeev John[3], Ajoy K. Kar[4], Roberto Osellame[1]

[1]Istituto di Fotonica e Nanotecnologie, Consiglio Nazionale delle Ricerche, Piazza Leonardo da Vinci, 32, I-20133 Milano, Italy.

[2]Laboratory of Artificial-Intelligence Nanophotonics, School of Science, Royal Melbourne Institute of Technology, GPO Box 2476, Melbourne, VIC 3001, Australia

[3]Department of Physics, University of Toronto, 60 St. George Street, Toronto, Ontario, Canada M5S 1A7

[4]Institute of Photonics and Quantum Sciences, David Brewster Building, Heriot-Watt University, Edinburgh, EH14 4AS, UK

*Correspondence to: arodenas@gmail.com

**Abstract:** Nanostructuring hard optical crystals has so far been exclusively feasible at their surface, as stress induced crack formation and propagation has rendered high precision volume processes ineffective. We show that the inner chemical etching reactivity of a crystal can be enhanced at the nanoscale by more than five orders of magnitude by means of direct laser writing. The process allows to produce cm-scale arbitrary three-dimensional nanostructures with 100 nm feature sizes inside large crystals in absence of brittle fracture. To showcase the unique potential of the technique, we fabricate photonic structures such as sub-wavelength diffraction gratings and nanostructured optical waveguides capable of sustaining sub-wavelength propagating modes inside yttrium aluminum garnet crystals. This technique could enable the transfer of concepts from nanophotonics to the fields of solid state lasers and crystal optics.

**Main Text:** The optical properties of a material are known to depend not only on its chemistry but also on its sub-wavelength structure. With the inception of the photonic crystal *(1, 2)* and metamaterial *(3, 4)* concepts this idea proved to be key to access a new level of light manipulation beyond what is allowed by the natural optical properties of materials. However, for over three decades of research no technique has been able to reliably nanostructure optical crystals beyond their surface.

Laser lithography as developed by the semiconductor industry is an intrinsic surface processing technique where a photoresist is two-dimensionally (2D) nanostructured with UV light, and, followed by various subtractive and additive processes, enables the mass production of high-quality nanophotonic planar devices *(5, 6)*. Its extension from 2D to 3D was demonstrated two decades ago by means of infrared femtosecond laser pulses to introduce multiphoton absorption at the resist photo-polymerization step *(7)*. This approach prompted seminal demonstrations of lithographically produced 3D photonic crystals for the optical range, first made of pure polymerized material *(8-10)*, and later transferred to silicon and other optical materials by means of further processing steps *(11, 12)*. The technological exploitation of photo-polymerized structures has however proven to be impractical since they cannot be efficiently interfaced with other photonic elements. A remarkable example of the potential of 3D nanostructuring of macroscopic optical materials comes from the field of optical fibers: photonic crystal fibers have delivered functionalities that go far beyond what is feasible with ordinary unstructured glass *(13, 14)* and have revolutionized the areas of nonlinear optics and optical communications *(15-17)*. Yet, their manufacturing in crystalline media has remained elusive due to the difficulty of applying stack and drawing techniques with crystals. The prospects of directly machining 3D nanostructures by laser-induced dielectric breakdown also led to reports that amorphous and void sub-micron structures can be induced in crystals, although at the cost of imparting high pressure waves which entail extended lattice damage and crack propagation *(18, 19)*. Despite efforts, no method has been reported for the large scale 3D volume nanostructuring of a crystal.

Departing from prevailing approaches, we propose that the inner chemical reactivity of a crystal, given by its wet etch rate, can be locally modified at the nanometer scale by means of multiphoton 3D laser writing (3DLW) in absence of amorphization and crack formation. We show that cm-long empty pore lattices with arbitrary feature sizes at the 100-nm level can be created inside some of the most used crystals in the laser industry (yttrium aluminum garnet and sapphire), in absence of brittle fracture. An etching selectivity larger than $10^5$, never observed before in a photo-irradiated material, is achieved between the modified and pristine crystalline states. This allows the design and fabrication of nanophotonic elements inside a crystal that provide optical responses determined by their sub-wavelength structure.

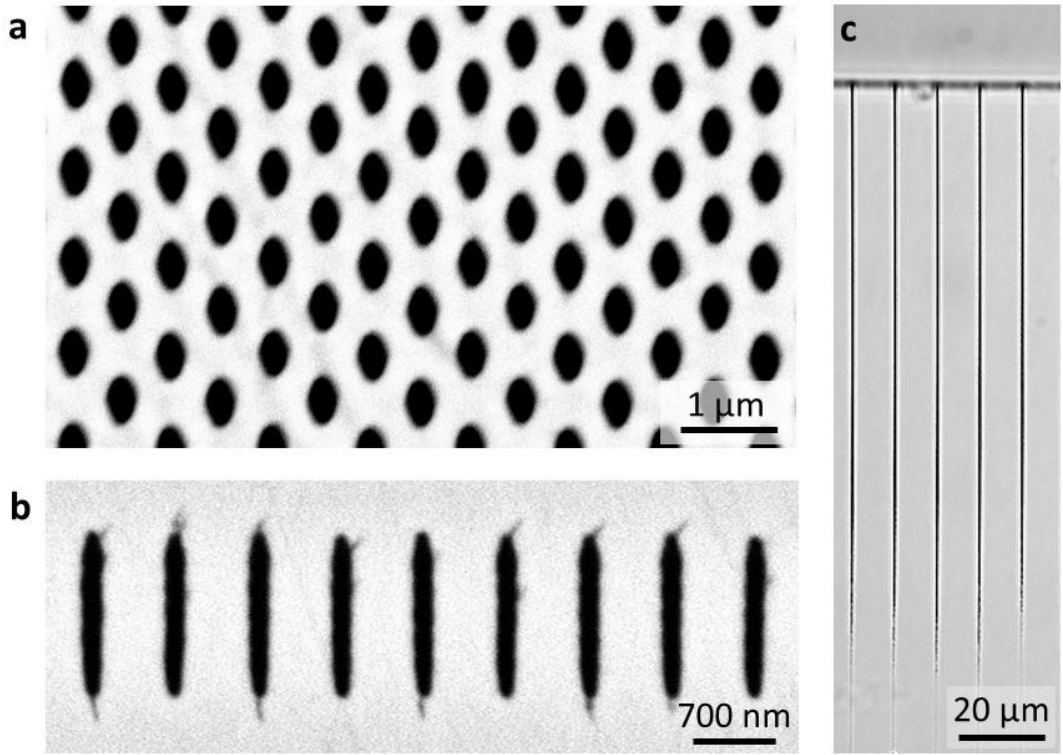

**Fig. 1. Wet etching of nanopore lattices fabricated by 3DLW in YAG.** (**a**) Nanopore lattice etched for 120h with average pore size of 257 nm ±7 nm and 454 nm ±13 nm along *x* and *y* reference axes, and 1 mm lengths along *z*. (**b**) Vertically overlapping nanopores after 2h wet etching (average size of 131 nm ±5 nm and 1300 nm ±35 nm, along *x* and *y*, and 1 mm lengths). (**c**) Top optical microscope view of 129 µm ± 6.8 µm long nanopores along *z* direction, etched for 1h.

To achieve volume nanostructuring of crystals at large scale it is necessary to arrange nanopores in arbitrary close-packed structures for macroscopic lengths (mm to cm scale) without causing brittle fracture of the crystal due to excessive stress accumulation. In addition, pores have to be written in arbitrary directions, their size must be controlled with at least ~10 nm reproducibility to achieve functional nanophotonic devices, and the pore cross-sectional shape must be tailorable.

We show that all these features (pore direction, size, shape, filling fraction, and length) can be controlled by combining 3DLW and wet etching of YAG crystal. Figure 1 summarizes the controlled creation of nanopore lattices in YAG crystals. No dependence on the crystalline axis was found for the pore size and wet etching rate within our experimental resolution *(20)*, so an arbitrary reference frame is set where *x* and *y* axes define the pore's cross sectional plane, and *z* is defined along the pore length axis. Figure 1A shows a 1 mm long lattice of nanopores arranged in centered orthogonal symmetry with 1 µm in-plane spacing. The lattice was etched for 120 h *(20)*, and the average pore size was of 257 ±7 nm and 454 ±13 nm in *x* and *y* directions, respectively. Control over the pore shape and size can be performed by tailoring the laser power and polarization (see Fig. S1), but to achieve bespoke pore shapes for constant laser power the overlapping of pores was also tested. Figure 1B shows an array of parallel pores which were vertically stacked so as to achieve arbitrarily elongated cross-sections. Pores in this experiment were etched for 2h and had an average size of 131 ±5 nm and 1300 ±35 nm along the *x* and *y* axes, respectively. The etching of crossing pores along different directions was also studied to prove the 3D nature of the lithographic process. Figure S2 shows pores that crisscross at 90º angles and at different relative depths within the crystal.

The creation of air pores photonic lattices makes it possible to achieve nanophotonic structures inside the crystal with a spatial resolution equivalent to state-of-the-art multiphoton polymerization lithography *(21)*. However, the use of nanophotonic devices in practical scenarios requires robust and efficient optical interconnections and capability for the design of large complex circuits. To achieve this, it is essential to keep the spatial resolution and lattice fidelity across large areas on the mm$^2$ scale or beyond. The critical parameter that limits such control over the pores lengths is the differential etching rate between the photo-modified written volumes and the surrounding pristine crystal. Figure 1C shows a top view optical microscope image of an array of 200 nm nanopores etched for 1h. The large refractive index contrast between air and crystal allow to determine a pore length of ~129 µm and thus an averaged nanopore etch rate of 129 ±6.8 µm/h for the first hour of etching. The slow etching rate of un-modified YAG was determined to be <1 nm/h (0.36 ±0.23 nm/h) from the average of all performed experiments in our wet etching conditions *(20)*. The 1h average etching selectivity of 3DLW YAG nanopores is thus determined to be greater than 10$^5$, the highest value ever observed for any lithographic process so far, and approximately two orders of magnitude higher than that of alumina over silicon *(22)*. Due to this high selectivity, nanopores with cross-sections of 368 x 726 nm$^2$ and lengths of 3.1 mm were achieved by etching for 170h from both end sides of the pores (see Fig. S3), showing that nanopores with mm-scale lengths are feasible in one single etching step. A similar ultrahigh selectivity and mm-scale nanopore length was also found for sapphire *(20)*. Furthermore, to homogeneously etch longer nanostructures on the cm scale (or potentially beyond), as well as to keep the etching times as short as a few hours, a scheme of vertical access etching pores was implemented, thus allowing the achievement of structures with arbitrary length across the whole sample (see Fig. S4).

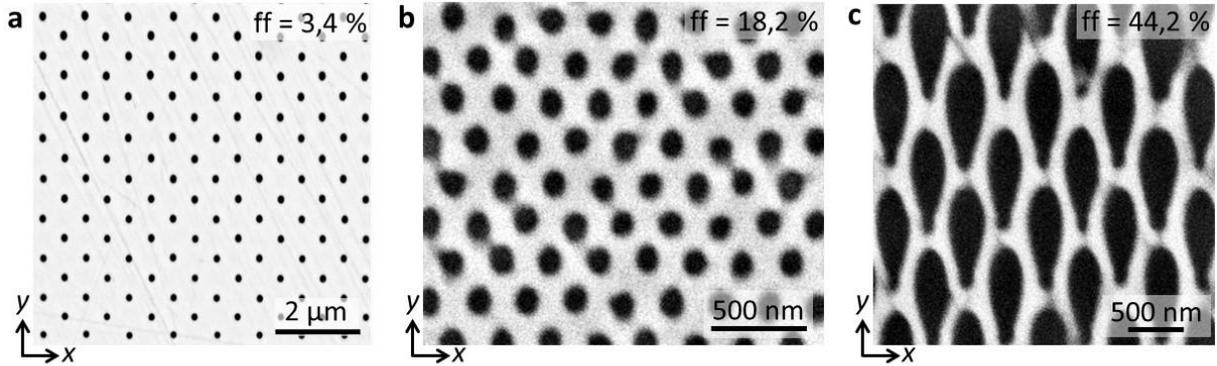

**Fig. 2. 3DLW nanostructures with 100 nm feature sizes and filling fraction control.** (a) Low pore filling fraction lattice (3,4%) written with linearly varying pore size along depth (top to bottom *x*-widths ranging from 139±4 nm to 109±6 nm, *y*-heights from 204±6 nm to 132±8 nm, and aspect ratio from ~1,45 to ~1,20). (b) Hexagonal pore lattice with 250 nm spacing and 18,2% pore filling fraction (average *x*-width 109 nm ±8 nm, *y*-height 125 nm ±10 nm, and aspect ratio ~1,15). (c) High filling fraction (44,2%) pore lattice with ~150 nm crystal walls separating elliptical pores with average 276 × 972 nm² cross sections. All pores had ~1 mm lengths.

As shown in Figure 1, the pores cross-sectional aspect ratio can be engineered by means of laser control and pore overlapping, however both these processes inevitably lead to larger pore cross-sections, and for some applications such as the fabrication of photonic crystals and metamaterials in the visible and IR ranges, symmetric pores at the 100 nm level are required. A route to minimizing both the pore size and cross-sectional aspect ratio (i.e. pore height along *y*-axis divided by width along *x*-axis) is by performing 3DLW at powers close to the threshold for laser photo-modification. Under our 3DLW experimental conditions *(20)* ~110 nm nanopores were obtained with almost circular shapes for circularly polarized laser irradiation (see Fig. S1). Figure 2A shows a lattice of 1 mm long nanopores fabricated with decreasing powers towards deeper layers. The obtained lattice has a slowly varying pore cross-section with widths ranging from ~139 nm to ~109 nm, heights from ~204 nm down to ~132 nm, and aspect ratios ranging from ~1.45 to ~1.20 at different depths. To test the feasibility of fabricating high filling fraction nanopore lattices without fracturing the crystal, hexagonal lattices with lattice spacing down to 250 nm were fabricated.

Figure 2B shows a close up of such lattice after etching for 15h, with an air filling fraction of 18%, average pore diameter of ~117 nm, aspect ratio of ~1.15, and a dielectric wall thickness of ~133 nm between pores. This control over the lattice spacing down to 250 nm could prove useful for designing 3D photonic band-gap lattices with stopbands ranging from the visible to the mid-IR range *(23)* inside solid state laser crystals such as rare earth ion doped YAG, one of the most used laser materials.

A different route to achieve larger air filling fractions is by etching larger pores. Large air filling fraction structures are also essential for tailoring the properties of microstructured optical waveguides (MOWs) such as the dispersion, mode size and nonlinear coefficient *(13-17)*. Reaching a high air filling fraction inside a crystal requires the fabrication of extended thin crystalline nanolayers without crystal fracture. The creation of such lattices was validated by fabricating a centered orthogonal lattice with in-plane pore spacing of 700 nm and pore cross-sections of 276x972 $nm^2$ (see Fig. 2C), having a 44% air pore filling fraction and ~150 nm dielectric walls in absence of crack formation even at the surface after mechanical polishing.

Having established the etching rates and achievable features sizes of the crystalline nanostructures inside YAG, we sought to determine whether: (1) the crystal nanopores are indeed void millimeters away from the entrant wet etching surface, and (2) functional sub-wavelength nanophotonics elements on the large scale can indeed be successfully realized. To test the first question, a nanograting with sub-wavelength pitch (700 nm) was fabricated to be tested at 1070 nm wavelength (see Fig. 3A). The sub-wavelength grating was designed to have only one -$1^{st}$ diffraction order (see inset in Fig. 3B), it was fabricated with 10 different laser powers, so as to control the pores cross-sectional shape, and tested in Littrow configuration for the highest diffraction efficiency *(24)*. The highest total efficiency of 86% was obtained for the -$1^{th}$ order of the grating that had the larger cross-section of the pores, 165 nm width and 1520 nm height. Calculations *(20)* assuming air filled pores are shown in Fig. 3B, agreeing well with the measured values and therefore demonstrating the void nature of the long etched nanopores.

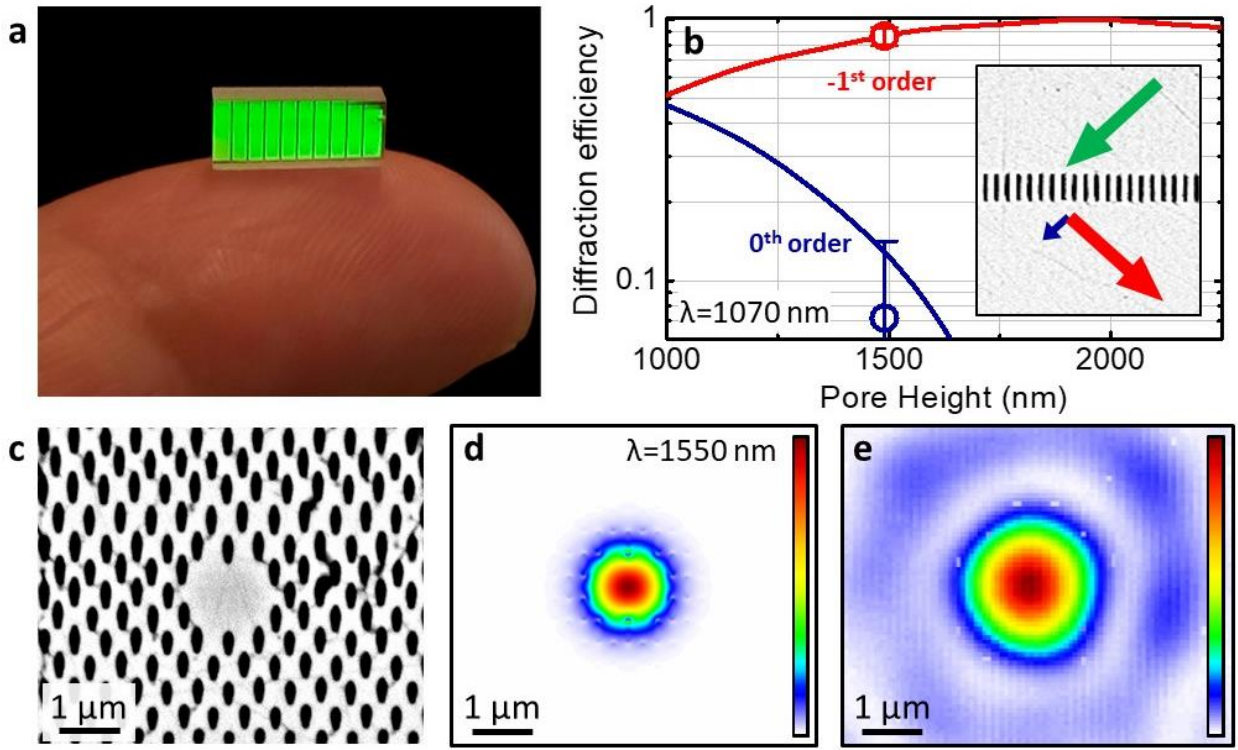

Fig. 3. **Sub-wavelength diffraction gratings and MOW (microstructured optical waveguide).** (**a**) Image of a cm-long 700 nm pitch grating under visible light illumination. (**b**) Experimental and calculated diffraction efficiency of a 700 nm pitch sub-wavelength grating for 1070 nm wavelength. Inset shows a SEM close up of the fabricated grating. (**c**) Optical waveguide with hexagonal structure, 500 nm horizontal pore-to-pore spacing, mean pore size of 166 x 386 nm$^2$ and 4 mm length. (**d**) Simulated intensity mode profile at 1550 nm with FWHM of 862 nm (vertical) and 972 nm (horizontal). (**e**) Diffraction-limited near-field image of the waveguide output mode measured at 1550 nm wavelength, with a FWHM of ~1,5 µm.

To further evaluate the quality and potential of the crystal 3D nanolithography technique, the fabrication of MOWs with different lattice spacings and cavity sizes was studied (see Fig. S5). Figure 3C shows a nanostructured waveguide with core size of 1.1 x 1.3 µm$^2$, a triangular symmetry cladding with in-plane spacing of 500 nm, an average pore size of 166 x 386 nm$^2$, and a length of 4 mm. Figure 3D shows the theoretical mode at 1550 nm wavelength for vertically polarized electric field, as calculated by means of finite element method (FEM). The mode is effectively sub-wavelength having a full width at half maximum (FWHM) of ~900 nm (~0.6λ). Figure 3E shows the measured diffraction-limited image of the waveguide output mode for vertical polarization at 1550 nm. From the FEM computations the MOW is expected to have a modified dispersion function with two theoretical zero dispersion wavelengths (ZDW) at 0.93 µm and 1.51 µm wavelength (see Fig. S6), well below the natural ZDW of YAG at 1.6 µm, which could allow for ultrafast nonlinear supercontinuum generation with standard commercial $Nd^{3+}$ and $Yb^{3+}$ short pulse lasers *(15)*.

Besides the novel applications on crystalline 3D nanophotonics and nonlinear optics, to which this technique gives a large impetus, the 3D structuring of YAG laser crystals also opens up new ways to overcome current limitations in the design of compact solid-state rare-earth-doped lasers, such as for the integration of traditional cavity elements (e.g. mirrors, dispersion control elements, microfluidic cooling channels) directly in the gain media itself with expected improvements in compactness, robustness and performance of the devices. The possibility of fabricating large nanostructured YAG laser crystals with reduced surface defects due to the wet etching process also opens up potential for new applications in ultra-strong deformable laser nanofibers *(25-27)*.

**Competing interests:** Non declared.

**Funding:** The work was funded by the European Union's Horizon 2020 research and innovation programme under the Marie Sklodowska-Curie Individual Fellowships EXTREMELIGHT project (Grant Agreement No. 747055). A. R. and R.O. acknowledge support from LASERLAB-EUROPE (Grant Agreement No. 654148, European Union's Horizon 2020 research and innovation programme). G. C. and R. O. acknowledge support from the European Research Council (ERC) Advanced Grant programme (CAPABLE, Grant Agreement No. 742745). M. G. acknowledges the support from the Australian Research Council (ARC) through the Discovery Project (DP170101775). S. J. acknowledges support from the US Dept. of Energy DOE-BES in a subcontract under Award No. DE-FG02-06ER46347. A. K. K. would like to acknowledge the UK Engineering and Physical Sciences Research Council (EP/M015130/1; EP/G037523/1).

**References and Notes:**

1. S. John, Strong localization of photons in certain disordered dielectric superlattices. *Phys. Rev. Lett.* **58**, 2486 (1987). doi: 10.1103/PhysRevLett.58.2486

2. E. Yablonovitch, Inhibited Spontaneous Emission in Solid-State Physics and Electronics. *Phys. Rev. Lett*. **58**, 2059 (1987). doi:10.1103/PhysRevLett.58.2059

3. R. A. Shelby, D. R. Smith, S. Schultz, Experimental verification of a negative index of refraction. *Science* **292**, 5514, 77-79 (2001). doi:10.1126/science.1058847

4. D. R. Smith, J. B. Pendry, M. C. K. Wiltshire, Metamaterials and negative refractive index. *Science* **305**, 5685, 788-792 (2004). doi: 10.1126/science.1096796

5. K. Jain, C. G. Willson, B.J. Lin, Ultrafast deep UV lithography with excimer lasers. *IEEE Electron Device Lett*. **3**, 3, 53-55 (1982). doi:10.1109/EDL.1982.25476

6. K. McGarvey-Lechable, T. Hamidfar, D. Patel, L. Xu, D. V. Plant, P. Bianucci, Slow light in mass-produced dispersion-engineered photonic crystal ring resonators. *Opt. Express* **25**, 3916-3926 (2017). doi:10.1364/OE.25.003916

7. S. Maruo, O. Nakamura, S. Kawata, Three-dimensional microfabrication with two-photon-absorbed photopolymerization. *Opt. Lett.* **22**, 132-134 (1997). doi:10.1364/OL.22.000132

8. M. Straub, M. Gu, Near-infrared photonic crystals with higher-order bandgaps generated by two-photon photopolymerization. *Opt. Lett.* **27**, 1824-1826 (2002). doi:10.1364/OL.27.001824

9. M. Deubel, G. von Freymann, M. Wegener, S. Pereira, K. Busch, C. M. Soukoulis, Direct laser writing of three-dimensional photonic-crystal templates for telecommunications. *Nat. Mater.* **3**, 444-447 (2004). doi:10.1038/nmat1155

10. S. Wong, M. Deubel, F. Pérez-Willard, S. John, G. A. Ozin, M. Wegener, G. von Freymann, Direct Laser Writing of Three- Dimensional Photonic Crystals with a Complete Photonic Bandgap in Chalcogenide Glasses. *Adv. Mater.* **18**, 3, 265-269 (2006). doi: 10.1002/adma.200501973

11. M. Hermatschweiler, A. Ledermann, G. A. Ozin, M. Wegener, G. von Freymann, Fabrication of silicon inverse woodpile photonic crystals. *Adv. Funct. Mater.* **17**, 14, 2273-2277 (2007). doi:10.1002/adfm.200601074

12. J. K. Gansel, M. Thiel, M. S. Rill, M. Decker, K. Bade, V. Saile, G. von Freymann, S. Linden, M. Wegener. Gold Helix Photonic Metamaterial as Broadband Circular Polarizer. *Science* **325**, 1513 (2009). doi:10.1126/science.1177031

13. J. C. Knight, T. A. Birks, P. St. J. Russell, D. M. Atkin, All-silica single-mode optical fiber with photonic crystal cladding. *Opt. Lett.* **21**, 1547-1549 (1996). doi:10.1364/OL.21.001547

14. P. Russell, Photonic Crystal Fibers. *Science* **299**, 5605, 358-363 (2003). doi:10.1126/science.1079280

15. J. M. Dudley, G. Genty, S. Coen. Supercontinuum generation in photonic crystal fiber. *Rev. Mod. Phys.* **78**, 1135–1184 (2006). doi:10.1103/RevModPhys.78.1135

16. J. G. Rarity, J. Fulconis, J. Duligall, W. J. Wadsworth, P. St. J. Russell, Photonic crystal fiber source of correlated photon pairs. *Opt. Express* **13**, 534-544 (2005). doi:10.1364/OPEX.13.000534

17. R. J. A. Francis-Jones, R. A. Hoggarth, P. J. Mosley, All-fiber multiplexed source of high-purity single photons. *Optica* **3**, 1270-1273 (2016). doi:10.1364/OPTICA.3.001270

18. E. N. Glezer, E. Mazur, Ultrafast-laser driven micro-explosions in transparent materials. *Appl. Phys. Lett.* **71**, 882 (1997). doi:10.1063/1.119677

19. S. Juodkazis, K. Nishimura, H. Misawa, T. Ebisui, R. Waki, S. Matsuo, T. Okada, Control over the crystalline state of sapphire. *Adv. Mater.* **18**, 11, 1361-1364 (2006). doi:10.1002/adma.200501837

20. Materials and methods are available in the Supplementary Information.

21. J. Fischer, J. B. Mueller, J. Kaschke, T. J. A. Wolf, A.-N. Unterreiner, M. Wegener, Three-dimensional multi-photon direct laser writing with variable repetition rate. *Opt. Express* **21**, 26244-26260 (2013). doi:10.1364/OE.21.026244

22. M. D. Henry, S. Walavalkar, A. Homyk, A. Scherer, Alumina etch masks for fabrication of high-aspect-ratio silicon micropillars and nanopillars. *Nanotechnology* **20**, 25 (2009). doi:10.1088/0957-4484/20/25/255305

23. J. D. Joannopoulos, P. R. Villeneuve, S. Fan, Photonic crystals: putting a new twist on light. *Nature*, **386**, 143-149 (1997). doi:10.1038/386143a0

24. T. Clausnitzer, T. Kämpfe, E.-B. Kley, A. Tünnermann, A. V. Tishchenko, O. Parriaux, Highly-dispersive dielectric transmission gratings with 100% diffraction efficiency. *Opt. Express* **16**, 5577-5584 (2008). doi:10.1364/OE.16.005577

25. L. R. Meza, S. Das, J. R. Greer, Strong, lightweight, and recoverable three-dimensional ceramic nanolattices. *Science* **345**, 6202, 1322-1326 (2014). doi:10.1126/science.1255908

26. J. Bauer, A. Schroer, R. Schwaiger, O. Kraft, Approaching theoretical strength in glassy carbon nanolattices. *Nat. Mater.* **15**, 438-443 (2016). doi:10.1038/nmat4561

27. A. Banerjee, D. Bernoulli, H. Zhang, M.-F. Yuen, J. Liu, J. Dong, F. Ding, J. Lu, M. Dao, W. Zhang, Y. Lu, S. Suresh. Ultralarge elastic deformation of nanoscale diamond. *Science* **360**, 6386, 300-302 (2018). doi:10.1126/science.aar4165

**Funding:** The work was funded by the European Union's Horizon 2020 research and innovation programme under the Marie Sklodowska-Curie Individual Fellowships EXTREMELIGHT project (Grant Agreement No. 747055). A. R. and R.O. acknowledge support from LASERLAB-EUROPE (Grant Agreement No. 654148, European Union's Horizon 2020 research and innovation programme). G. C. and R. O. acknowledge support from the European Research Council (ERC) Advanced Grant programme (CAPABLE, Grant Agreement No. 742745). M. G.

acknowledges the support from the Australian Research Council (ARC) through the Discovery Project (DP170101775). S. J. acknowledges support from the US Dept. of Energy DOE-BES in a subcontract under Award No. DE-FG02-06ER46347. A. K. K. would like to acknowledge the UK Engineering and Physical Sciences Research Council (EP/M015130/1; EP/G037523/1).

**Author contributions:** Investigation: A. R.; Conceptualization: A. R., M. G., and R. O.; Methodology: A. R., G. C., P. P. and R. O.; Validation: A. R., P. P. and G. C.; Formal Analysis: A. R., S. J.; Resources: A. R. and R. O.; Writing Original: A. R. and R. O.; Writing Review: All authors; Visualization: A. R. and G. C.; Supervision: A. R., M. G., S. J., A. K. K. and R. O.; ([CRediT](#) taxonomy).

**Competing interests:** Non declared.

**Data and materials availability:** All data needed to evaluate the conclusions is available in the main text or the supplementary materials.
14

# SUPPLEMENTARY INFORMATION

## Three-dimensional laser nanolithography of laser crystals

Airán Ródenas[1*], Min Gu[2], Giacomo Corrielli[1], Petra Paiè[1], Sajeev John[3], Ajoy K. Kar[4], Roberto Osellame[1]

*Correspondence to: arodenas@gmail.com

**This PDF file includes:**

Materials and Methods
Supplementary Text
Figs. S1 to S7
References

**Materials and Methods**

Sample fabrication

Commercial yttrium aluminum garnet (YAG) (FOCtek Photonics, Inc) with sizes 20x10x2 mm$^3$ and crystal orientation <111>, and c-plane <0001> sapphire substrates, were processed by means of standard three-dimensional laser writing (3DLW) with an ytterbium mode-locked ultrafast fiber laser with 1030 nm wavelength, 350 fs pulse duration, and linear polarization (Satsuma, Amplitude Systemes). The laser repetition rate for all the experiments here presented was set at 500 kHz, however trials at 1 MHz were found to yield qualitatively equal results, not shown here for the sake of brevity. A half-wave plate in combination with a linear polarizer were used to control the laser power, and a quarter-wave plate was used to convert linear polarization to circular. A 1.4 numerical aperture (NA) oil-immersion Olympus objective was used to tightly focus the laser pulses inside the crystals. The working distance of the used objective was of 0.15 mm which limits to around 0.2 mm the maximum depth at which the material can be processed considering the refractive index of the crystal. Three dimensional nanopositioning of the sample was done by means of computer controlled XYZ linear stages (ANT-series, Aerotech, Inc). Line scanning speeds were always of 1 mm/s (2 mm/s in the case of 1 MHz repetition rate).

After laser irradiation, crystals were laterally polished to expose irradiated structures and perform wet chemical etching. YAG crystals were etched in phosphoric acid 44 wt.% solution in deionized water. Etching was performed in a magnetic stirrer with a digital ceramic heating plate at 350 K (IKA C-MAG HS 4). Sapphire crystals were etched in 20 wt.% HF solution in deionized water, at 308 K in ultrasonic bath. After etching, samples were cleaned in three consecutive ultrasonic baths with deionized water, acetone, and methanol, respectively. In the case of YAG, no dependence on the crystalline axis was found for the pore size and wet etching rate within our experimental resolution, so an arbitrary reference frame was set where $x$ and $y$ axes define the pore's cross-sectional plane, and $z$ defines the pore length axis. Writing laser direction was along $-y$ axis in all figures. In the case of sapphire, only one laser fabrication experiment was performed with pores within the <0001> c-plane at an arbitrary direction.

Sample characterization

Back-scattered scanning electron microscopy (BSD-SEM) was performed with a compact tabletop SEM on uncoated crystals at 10 kV (Phenom Pro, Phenom World). All images of air pore lattices were digitally processed with the open-source Fiji image processing package *(28)* to obtain all statistical size and shape distributions.

Fabricated sub-wavelength diffraction gratings were characterized at 1070 nm with a fiber laser. A linear polarizer was used to set the polarization to transversal electric (TE) with respect to the grating and a photodiode (PD300, OPHIR) was used to measure the power of the incoming, reflected, zero order and -1 diffracted order beams, respectively. Microstructured optical waveguides (MOWs) were characterized at 1310 nm and 1550 nm with laser diodes and linear polarizers, and the imaging of the MOWs output modes was performed on an InGaAs focal plane array camera with 640x512 pixels (Bobcat-640-GigE, Xenics). Free space light in- and out-coupling from the MOWs was performed with 0.68 and 0.55 NA aspheric lenses respectively, which had anti-reflection coating (1050-1600nm) (C330TM-C, Thorlabs).

Theoretical calculations and simulation methods

The theoretical design of the sub-wavelength gratings was performed following the two-mode interference model developed by Clausnitzer et al. *(24)* for loss-less dielectric transmission gratings. The effective refractive indices of the grating were numerically simulated using the finite element method (FEM) in COMSOL Multiphysics 4.2. Perfectly-matched-layers were applied outside the computed regions to obtain open boundaries. The refractive index dispersion and Sellmeier coefficients of undoped YAG from the UV to mid-IR ranges were taken from Zelmon et al. *(29)*. Modelling of the YAG MOWs was also performed by the same FEM software and method.

**Supplementary Text**
3DLW nanolithography of YAG crystals

Nanostructuring of transparent solids by means of laser techniques has been a topic of research for many decades. A relevant example has been the fabrication of embedded nanograting structures in glass, first demonstrated by Shimotsuma et al. *(30)*. In this case, nanostructuring is achieved only along one axis and not in 3D (the modification has micrometer size in the other two axes). In addition, the structures are strictly periodical and therefore cannot be used for arbitrary 3D nanostructuring. Possibly, the only technique hitherto capable of achieving fully arbitrary 3D nanostructures with ~100 nm features sizes is multiphoton polymerization (MPP) *(7-12)*. In this technique, photosensitive resins are used as initial materials, and photopolymerization is typically induced by nonlinear absorption of focused femtosecond laser pulses *(21)*. Translation of the sample along arbitrary trajectories allows the fabrication of complex structures by 3D laser writing (3DLW). The resulting 3D photonic structures are characteristically isolated in space, have micrometric size, need supporting walls and suffer from shrinkage and low damage threshold. In contrast to MPP, the technique presented here allows to directly nanostructure an optical crystal, maintaining the characteristic 3D spatial resolution and design freedom of 3DLW, while enabling the fabrication of large photonic structures that, being embedded in the volume of an optical crystal, can be easily coupled to other external or internal optical components. The laser irradiation procedure is equal to that in standard MPP, i.e. 3DLW, where tightly focused sub-picosecond laser pulses photo-modify the crystalline lattice at the nanoscale and within its volume, creating a single voxel of modified material which is the building block to write arbitrary 3D nanostructures.

Previous works exploiting 3DLW in YAG crystals to achieve 3D photonic structures where based on the micro-explosion method *(31)* (i.e. optical breakdown) and resulting structures had excess stress, often cracked if the in-plane pore spacing was lower than 2 µm and had vanishing photonic properties due to the very low refractive index contrast between photo-modified and un-irradiated volumes. The nonlinear photo-modification process in femtosecond-pulse 3DLW in YAG crystals has recently been assigned to a multiphoton ionization process corresponding to a five-photon absorption *(32)*, which explains the nanometric resolution that is achieved in the present work. The creation of micrometric refractive index increased regions within YAG crystals by means of 3DLW below the optical breakdown regime and in absence of apparent amorphization has been demonstrated, which allows to fabricate step-index optical waveguide circuits embedded in the crystal *(33)*. The fabrication of large microfluidic channels within YAG was also demonstrated by etching photo-modified YAG regions below the optical breakdown regime *(34)*, and the enhancement in the crystal etching rate was tentatively assigned to the creation of distorted YAG lattice, by means of a micro-Raman study of the large modified volumes *(34)*.

In this work, the laser power dependence of the diameter of etched nanopores was studied for both linear and circular laser beam polarizations. Circular polarization was always used for fabrication of photonic structures (such as gratings or microstructured optical waveguides) since that polarization state allows to reproducibly fabricate air pores in the nanoscale region below 200 nm with close to circular cross-section, as well as because it decouples the laser scanning and electric field polarization directions. The laser power was measured before the focusing lens and the transparency of the lens at 1030 nm (45%) was taken into account to obtain an estimate of the laser power at the focus. Figure S1(A) shows the pore cross-sectional width and height dependence on laser power. Although for both polarizations pore widths of 120 nm can be produced, we found a surprisingly low height cross-section only for circular polarization. The aspect ratio of the pores, calculated as the height to width ratio, is shown in Fig. S1(B). As in standard MPP laser lithography *(21)*, the cross-sectional aspect ratio for linear polarization is found to range between ~3.1 and ~3.75. This aspect ratio is a known limitation in standard MPP lithography which implies that for obtaining symmetric lattices along the vertical and horizontal directions, a multiscan approach must be followed to thicken up the polymerized lines. This limitation implies that in practice the spatial resolution in standard MPP is reduced by a factor of ~3.5, given by the minimal line heights achievable. In contrast to this, as shown in Fig. S1(B), pores with aspect ratios close to 1 can be achieved in YAG, which allows to design photonic lattices with symmetric profiles in the 3D space with simultaneously spatial feature sizes in the 100 nm scale.

Multiscan in 3DLW of solids, i.e. the possibility to write closely spaced features, requires that written features do not scatter light excessively. In the case here reported, material is not removed with the irradiation but a faint modification is performed that appears to not affect significantly subsequent irradiations. Fig. S2A shows an image of a partially etched pore obtained in bright-field transmission microscopy. The un-etched region is barely visible with respect to the etched one, because in the former case the index change is very low as compared to that achieved in the etched portion, where, although of similar dimensions, the track is filled with air. Precise 3D pore overlapping capability has also been observed: in Fig. 1b for example, we show that different tracks can be overlapped with high precision along the in-depth writing direction. The etching of pores along arbitrary directions within the crystal was also studied to prove the 3D capability of the etching process. Figure S2B shows a 3D scheme of crossing pores at different relative depths and 90º angles. SEM pictures of perpendicular cuts (made by mechanical polishing) of equally crossing pores are shown in Fig. S2(C, D). These experiments were done at high laser power so that characteristic lobes, induced by spherical aberration to the focused laser beam, are transferred to the photo-modified and etched volumes and are thus visible in the figure. The observed surface roughness corresponds to sputtered silver nanoparticles created in the metal coating process before the SEM imaging. A close-up view of the inside of one of the slanted pores is shown in Fig. S2E, where a smooth inner surface of the pores can be observed, which may indicate that the roughness is below or equal to the resolution of the used SEM system (around 3 nm). Other air pore directions along horizontal and vertical axis are shown in Fig. S4.

Wet etching of 3DLW nanostructures inside garnet crystals.

Wet etching of garnets and other oxide crystals in hot phosphoric acid is well known since the 60's, see for example *(35)*. The chemical reaction which may explain the wet dissolution of YAG has not been reported in the literature. A possible reaction could be:

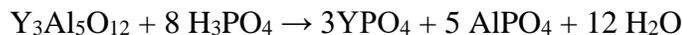
$$Y_3Al_5O_{12} + 8\ H_3PO_4 \rightarrow 3YPO_4 + 5\ AlPO_4 + 12\ H_2O$$

At this stage, we do not know which exact mechanism produces the observed etching selectivity at the molecular level within photo-modified and pristine YAG. However, due to the fact that very dense pore lattices can be fabricated in absence of brittle fracture of samples (see

Fig. 2), we ascribe the ultrahigh etching selectivity between pristine and photo-modified regions to the creation of lattice defects at the laser ionized volumes in the absence of lattice phase changes to an amorphous YAG phase, which would entail crack propagation.

Since other garnets than YAG are known to etch well in $H_3PO_4$ acid *(35)*, we expect that the present lithographic technique could be valid also for them. The use of yttrium iron garnet (YIG) crystals would be particularly interesting in photonic applications due to its higher index than YAG (2.20 for YIG and 1.81 for YAG, at 1550 nm wavelength), as well as due to its well-known magneto-optical properties. Extension of the technique to other optical crystals such as sapphire was tested with similar ultra-high selectivity results (see below).

Nanopores with mm-scale lengths.

To achieve practical devices, nanophotonic lattices should have footprints from the micrometric to the cm scale. This implies that pores should have length to diameter ratios on the order of $10^5$, a feature currently unachievable by any fabrication technique in brittle materials.

We tested the etching of long pores on the sub-cm scale: long (3.1 mm along *z* reference axis) nanopores were fabricated with both facets polished so that wet etching could evolve from both sides. Pores had cross-sections of 368 x 726 $nm^2$ and were separated by 10 µm so that they can be optically distinguished under microscope inspection. Figure S3(B) shows a composed microscope image of the etched pores after around 170 h. Pores were written with circular polarization and laser powers ranging from 6.5 to 7 mW. Figure S3(C) shows the SEM close-up of the etched nanopores.

Nanopores with arbitrarily long lengths.

The main limitation in the etching length of 3DLW lines is the difficulty in refreshing the exhausted acid inside the nanopores. When lattices need to have footprints beyond the mm scale, a strategy for etching arbitrarily long structures was developed profiting from the 3D capability of the lithography method: auxiliary vertical etching pores (VEPs) were designed and successfully implemented to provide additional access points for the acid beside the end facets, or even without them. This approach allows to fabricate complex 3D nanostructures with arbitrary lengths. Figure S4(A) shows a 3D sketch of such VEPs in red as implemented for etching out infinitely long MOW structures. This scheme was exploited to fabricate all MOWs and large gratings. Figure S4(B) shows the SEM image of a polished MOW which partially reveals a VEPs structure cut through. A top optical view of an array of 8 mm long MOWs where VEPs were evenly distributed every 80 µm is shown in Fig. S4(C). This etching architecture allows to etch nanostructures evenly at all required points and also shows the 3D capability of the 3D nanolithography technique.

Microstructured optical waveguides (MOWs) in YAG crystals.

The theoretical properties of fabricated MOWs were evaluated through FEM computation. The effective indices of the fundamental modes of all MOWs were computed from the UV region (250 nm) to the mid-IR (5500nm). From this data the wavelength dependent dispersion parameter of the MOWs was evaluated, obtaining similar results to those observed in microstructured optical fibers (MOFs) and photonic crystal fibers (PCFs) made of glass. The appearance of two zero-dispersion wavelengths for MOWs with micron and sub-micron lattice parameters is found to be a characteristic of these waveguides (see Fig. S6), which could prove useful in supercontinuum coherent light generation with standard 1-µm wavelength pump lasers.

The transmission properties of MOWs were characterized experimentally with free space optics in the near-IR range. A variety of different MOWs with in plane lattice spacings of 500-700 nm and different cavity sizes were also characterized. Almost all of them supported guided modes at both 1310 nm and 1550 nm wavelengths, as shown in Fig. S5, for TM vertical polarization. The guided modes are wavelength to sub-wavelength in FWHM size, beyond the resolution limit of the used imaging system that was limited by the numerical aperture of the used lens (NA=0.55). This resolution was experimentally characterized at 1.55 µm wavelength with a resolution target, to be of around 358 lp/mm, that is, capable of resolving features no smaller than ~1.4 µm. We believe that for this reason, all the measured modes have apparently similar FWHMs, even though simulations of mode size (see Fig. 3d) predict sub-wavelength size. Finally, we note that all damaged regions in Fig. S5(C-E) were due to mechanical polishing of the crystal and were only superficial.

The losses of the 3DLW fabricated MOWs featuring mode sizes on the sub-wavelength scale, were measured to have values on the dB/mm level. For the case of the highly confining waveguide with a sub-wavelength mode shown in Fig. 3c, which has a calculated mode FWHM of $\sim 0.6\lambda$ at 1.55 µm wavelength, we measured an insertion loss of 36 dB in the used free-space coupling setup. We calculate a loss of ~7 dB due to mode mismatch on each facet between the waveguide mode and the free space focused spot (0.55 NA lenses). We also estimate ~0.3 dB due to reflection losses on each facet, plus losses due to misalignments between the waveguides and the lenses axes, which we are not able to estimate precisely. All these contributions sum up to a minimum of coupling losses of ~15 dB. This yielding an upper limit for propagation losses of ~21 dB for a 4 mm long waveguide, i.e. a propagation loss of ~5 dB/mm. The measured losses could probably be substantially improved by engineering the waveguide cladding microstructure design and the pore cross-section, both factors influencing effective index contrast and field exposure to scattering surfaces. And additionally, by optimizing thermal annealing treatments, modifying the 3DLW parameters such as pulse repetition rate and duration, and the wet etching solution. Moreover, utilizing the 3DLW capability for 3D design prototyping, coupling losses could be greatly reduced by designing mode tapers that adiabatically increase the mode size to match that of commercial fibers.

Nanopore fabrication in sapphire crystals

Laser structuring of sapphire crystals has been a subject of research since more than a decade now. Important seminal work was done by Juodkazis et al *(19)*, which demonstrated the creation of microfluidic channels. To evaluate if the observed giant nanopore etching selectivity is not restricted to the case of YAG, but it is a phenomenon that can also manifest itself in other types of crystals, we performed preliminary nanostructuring experiments in sapphire.

Under the same 3DLW conditions as in YAG and circular laser polarization, nanopores of ~200 nm cross-section and mm-scale lengths were observed to form in c-axis cut sapphire substrates (sere Fig. S7). Nanopores were observed to form from a minimum laser power threshold of ~4 mW (significantly lower than the observed threshold for YAG of ~7 mW; see Fig. S1A), for which an average width of 121 nm ±12 nm and height of 135 nm ±21 nm was measured, confirming than 3D nanolithography at the 100 level can also be achieved in sapphire (see Fig. S7C).

In contrast with YAG, in sapphire the appearance of multiple pores along the writing direction was observed, which we ascribe to birefringence-induced focus splitting in sapphire in contrast to the case of YAG for which this aberration is absent due to its index isotropy. Birefringence-induced focus splitting in 3DLW of crystals has been previously studied, and occurs when focusing is performed at high numerical aperture along the optical c-axis of the birefringent crystal *(36)*. We also disregard a self-focusing mechanism being at the origin of the observed focus splitting, due to the low pulse peak powers used in this technique (~$5 \times 10^{-2}$ MW) in comparison with the critical self-focusing power of ~3 MW for sapphire ($P_{cr}=0.15\lambda^2/n \cdot n_2$, with n ~1.74 and $n_2$ ~$3 \times 10^{-16}$ cm$^2$/W).

To etch laser-written sapphire crystals, both $H_3PO_4$ and HF acid solutions were tested and only the latter produced observable etching of pores. The etching rate of nanopores was observed to be maximal for pores written at ~10 mW, for which an etching rate of 78.8 ±0.7 µm/h was measured during the first hour of etching. The etching rate of un-modified sapphire was determined by measuring the difference in width and height of different pores etched for 40 hours, and determined to be of <1 nm/h (0.34 ±0.25 nm/h) in average. These different etching rates yield a selectivity of ~$10^5$ (2.3 ±1.7 x$10^5$), similar to that found for nanopores in YAG (3.6 ±2.5 x$10^5$), and more than one order of magnitude larger than that previously observed for etching amorphous microchannels in sapphire *(19)*.

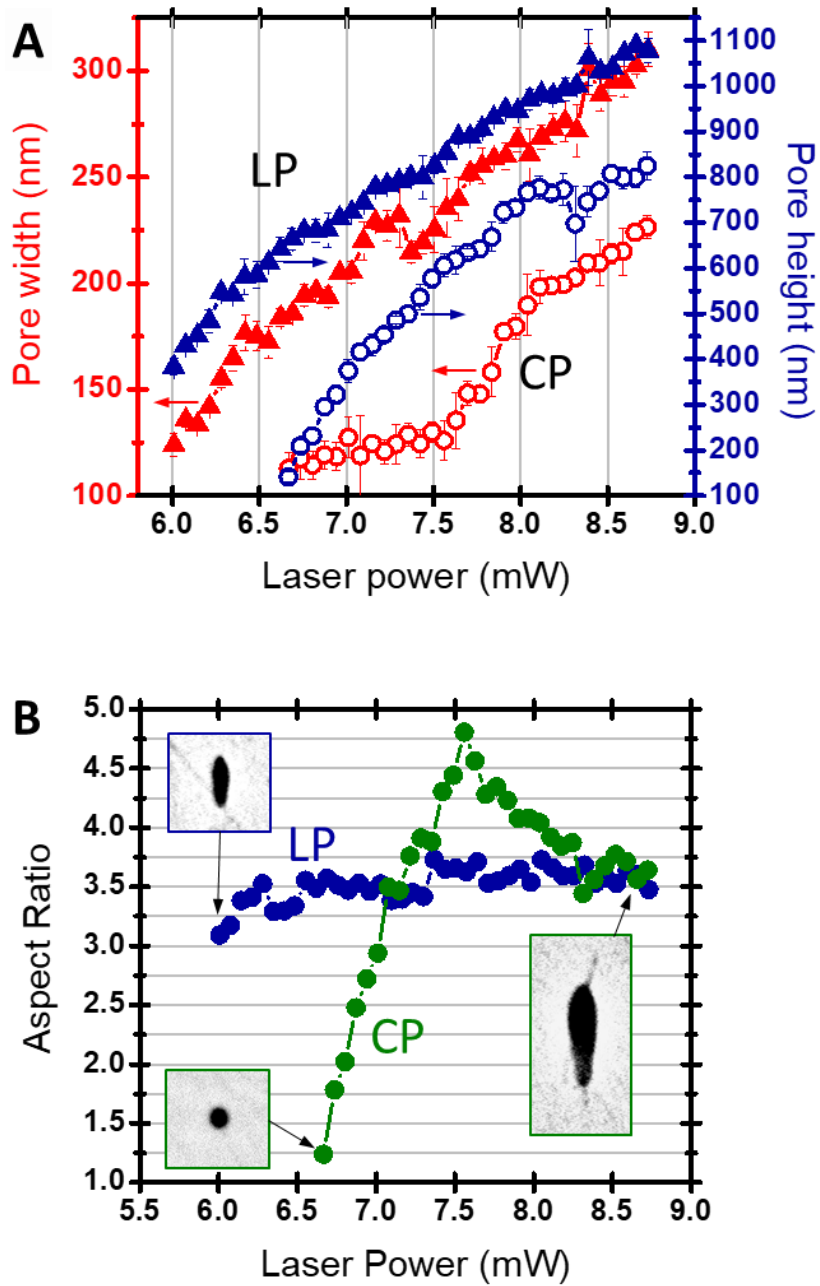

**Fig. S1. Evolution of pore size and cross-sectional aspect ratio as a function of laser power for linear and circular polarizations.**

(**A**) Power dependence of pore widths (in red) and heights (in blue) for linear (LP) and circular (CP) polarizations, measured from pores etched for 1h. (**B**) Dependence of cross-sectional pore aspect ratio (height divided by width) for linear and circular polarizations.

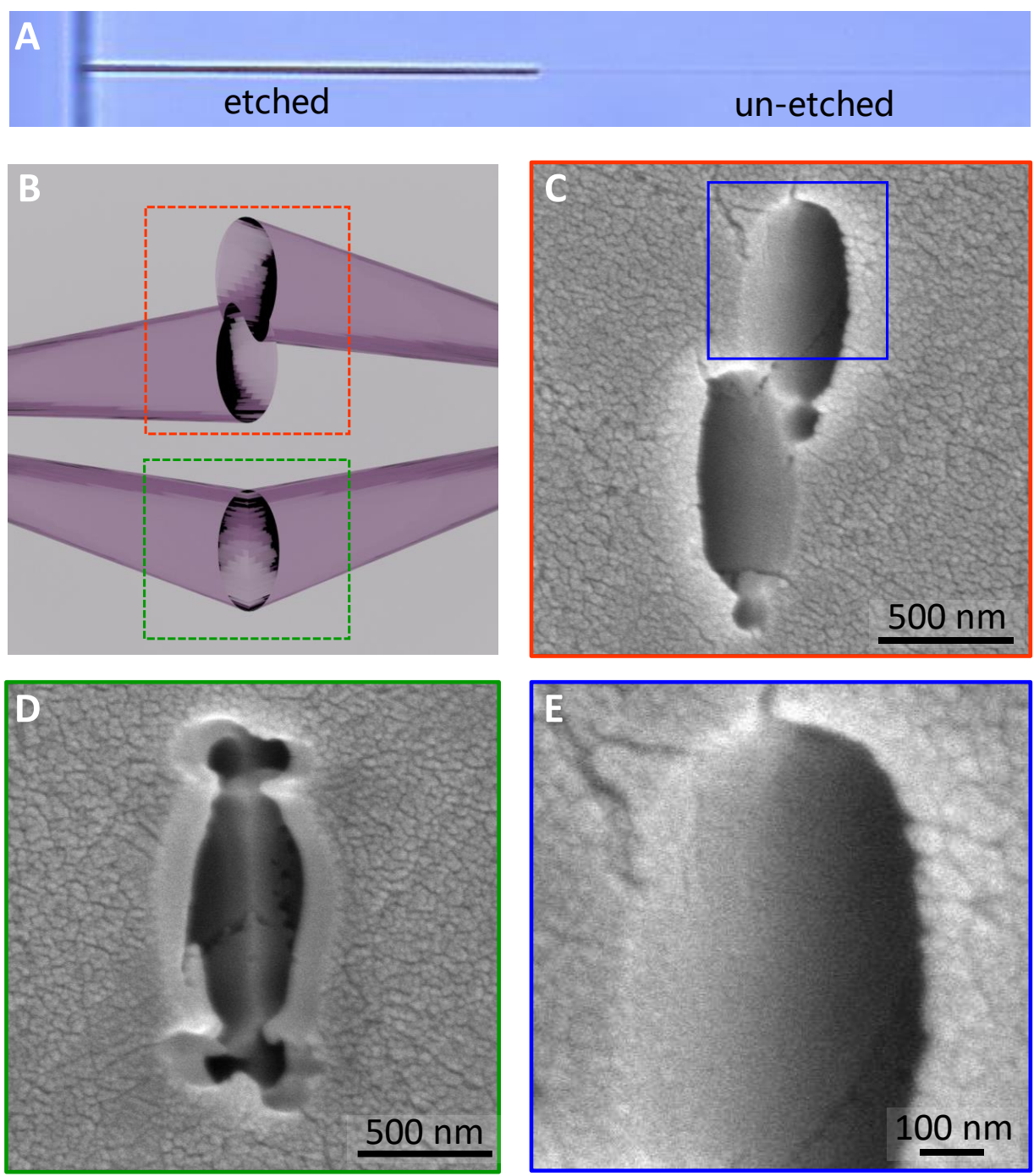

**Fig. S2. Etching of criss-crossing nanopores.**
(**A**) The large index contrast between etched and un-etched pores is depicted in a raw bright-field transmission image. (**B**) 3D sketch of 90º crossing pores at different vertical offset positions. (**C-D**) SEM pictures of crossing pores at 90º and different crossing heights. Ag sputtered nanoparticles are also visible on the main surface. (**E**) Close-up view of the inner smooth surface of a pore.

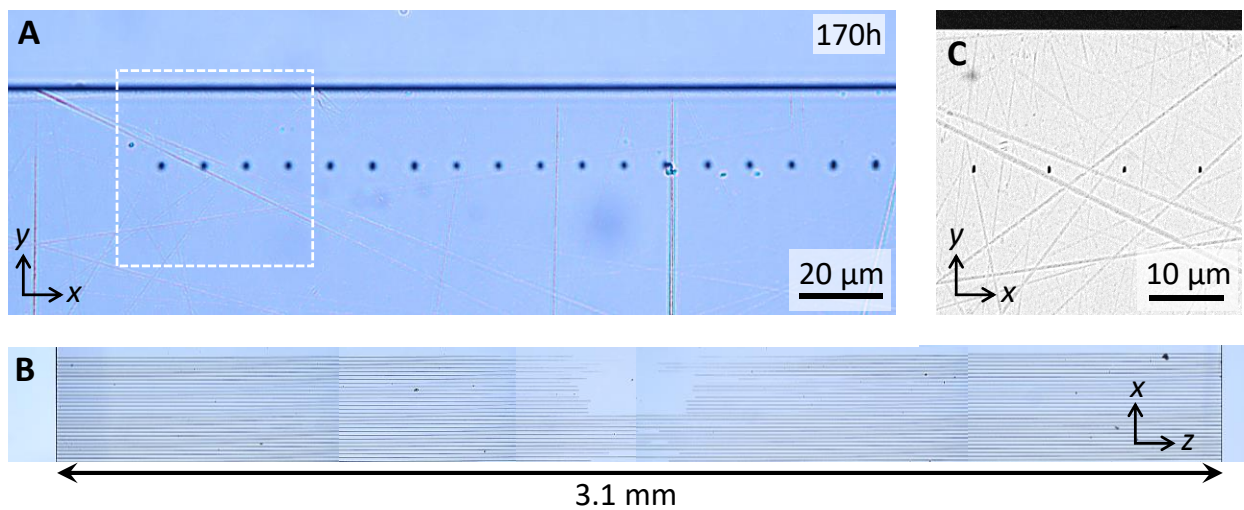

**Fig. S3. Etching of nanopores along mm to cm scale lengths.**
(**A**) Optical microscope side view of etched pores. (**B**) Optical microscope top view of etched nanopores. (**C**) SEM side view of etched nanopores.

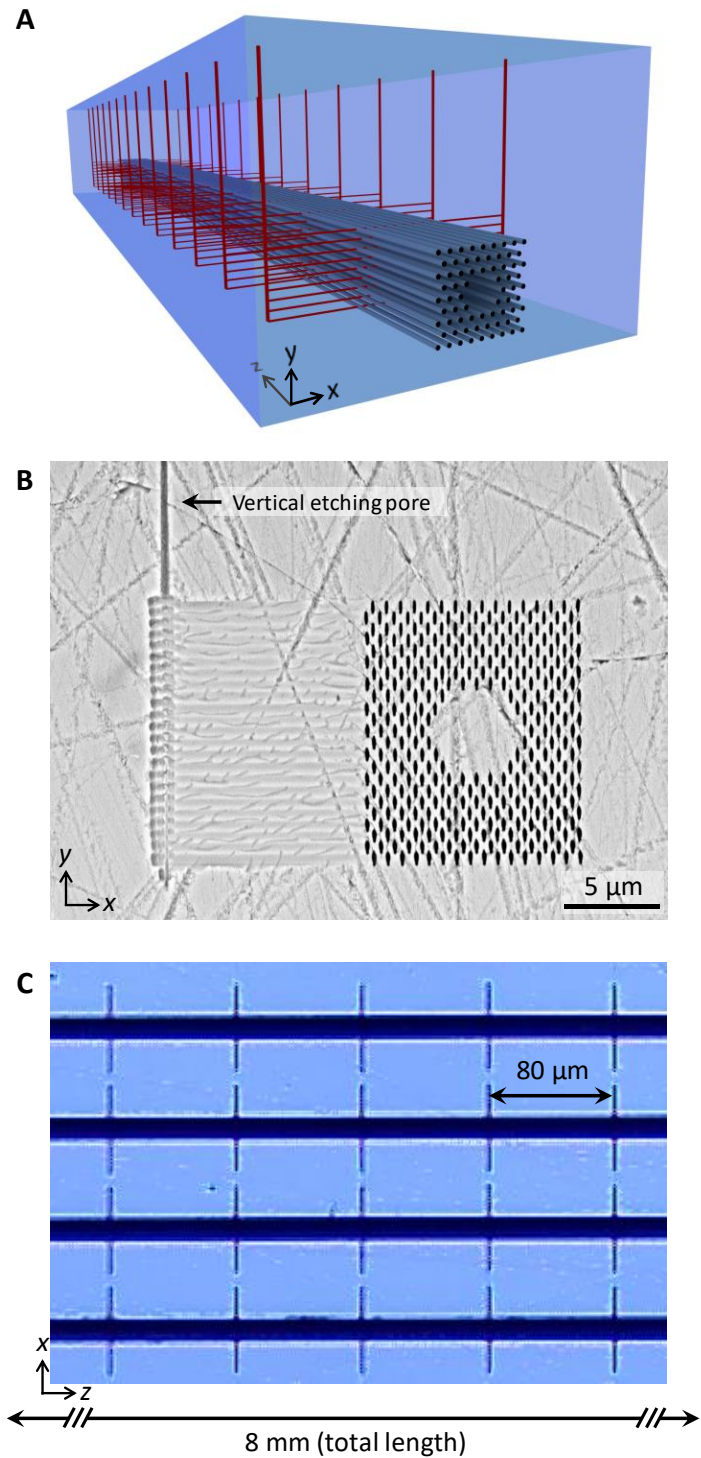

**Fig. S4. Achieving infinitely long and homogeneously etched nanopore lattices by means of 3D-connecting etching pores.**
(**A**) 3D sketch of the vertical etching channels architecture for etching MOWs. (**B**) SEM of a polished cut through a MOW partially revealing 3D etching pores. (**C**) Microscope top view of an etched array of MOWs with vertical etching channels every 80 µm.

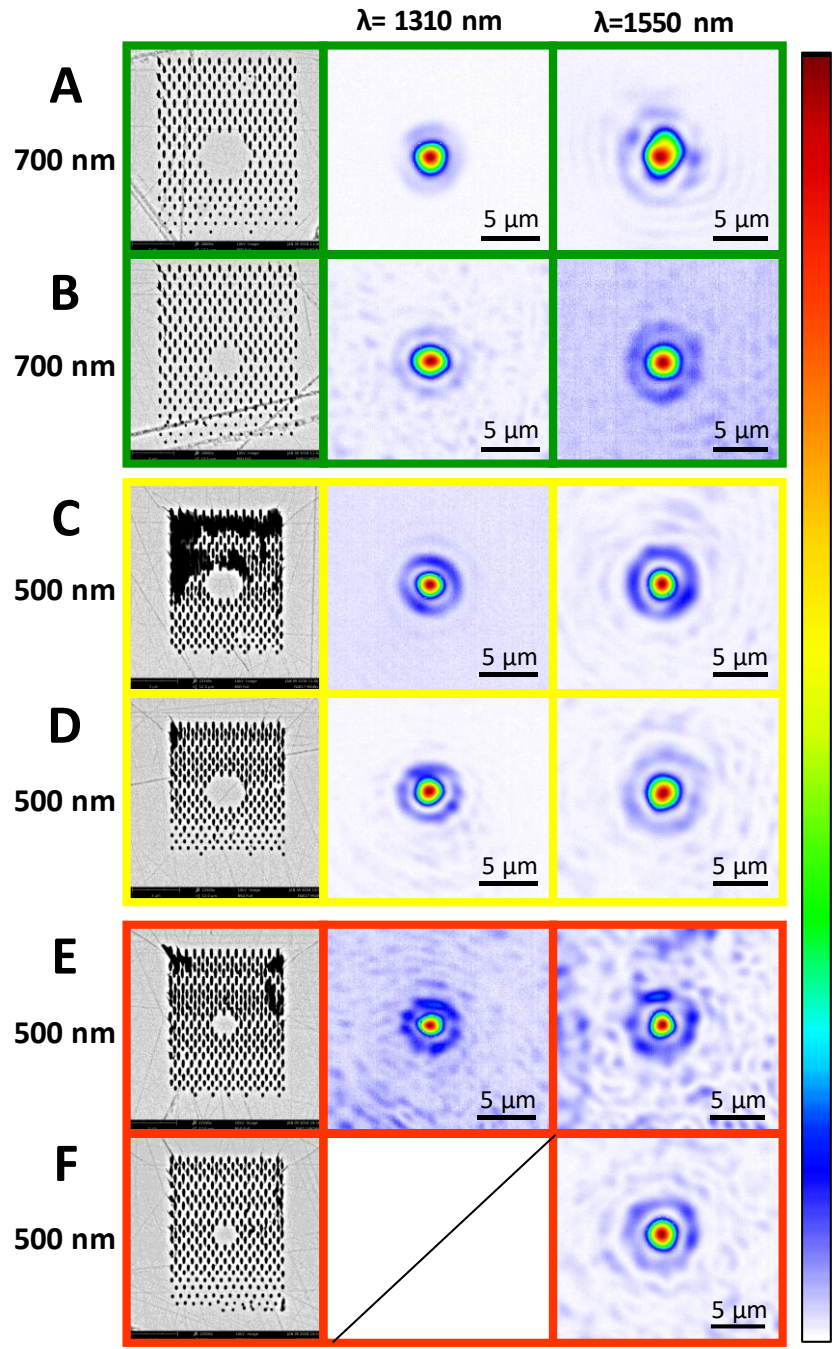

**Fig. S5. Optical characterization of output modes of different MOWs with 4 mm length.**
(**A**) and (**B**) MOWs with different cavity size and in plane spacing of 700 nm. (**C**) and (**D**) MOWs with wavelength-size cavity and in-plane spacing of 500 nm. (**E**) and (**F**) MOWs with sub-wavelength cavity size and in-plane spacing of 500 nm. All images are normalized to the color scale shown and correspond to vertical TM polarization. Broken facets of the MOWs are due to mechanical polishing of the crystal and are only superficial.

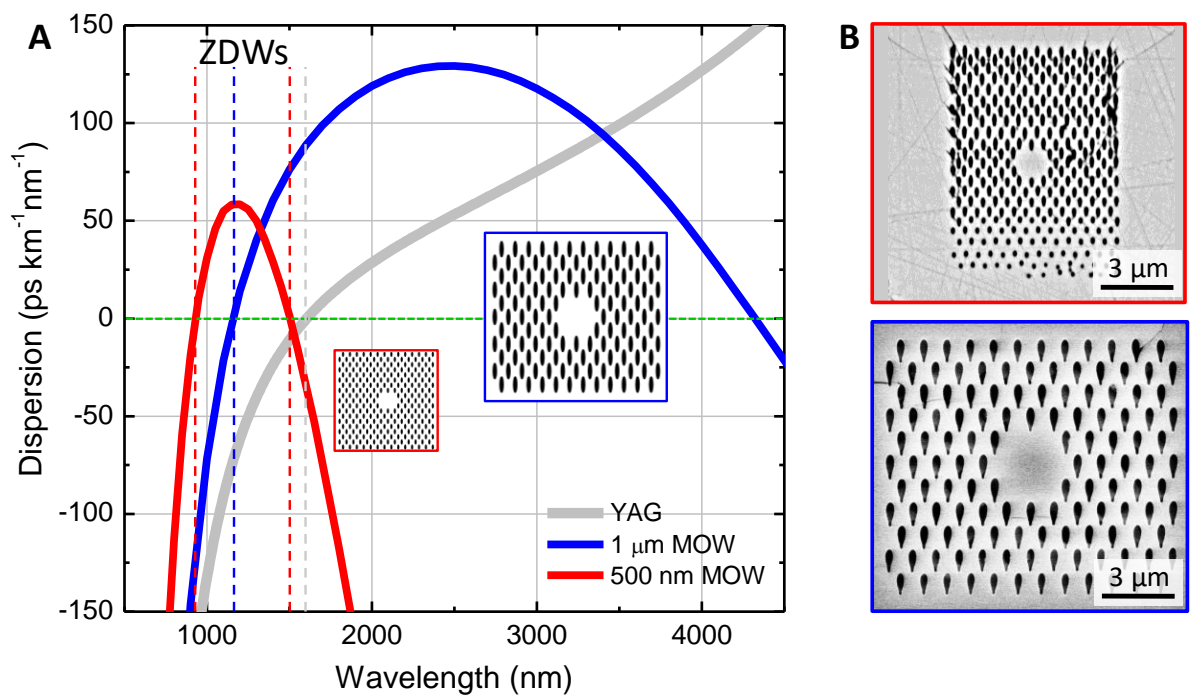

**Fig. S6. Dispersion control of 3DLW MOWs (microstructured optical waveguides).**
(**A**) Calculated wavelength-dependent dispersion of two different MOWs with horizontal lattice spacings of 500 nm (red) and 1 µm (blue). The dispersion of bulk YAG crystals is also shown. The hexagonal lattice MOW (red) has two zero dispersion wavelengths (ZDW) at 933 nm and 1509 nm. The orthogonal lattice MOW (blue) also has two ZDWs at 1164 nm and 4330 nm. (**B**) SEM images of fabricated MOWs corresponding to calculated dispersion curves in (A).

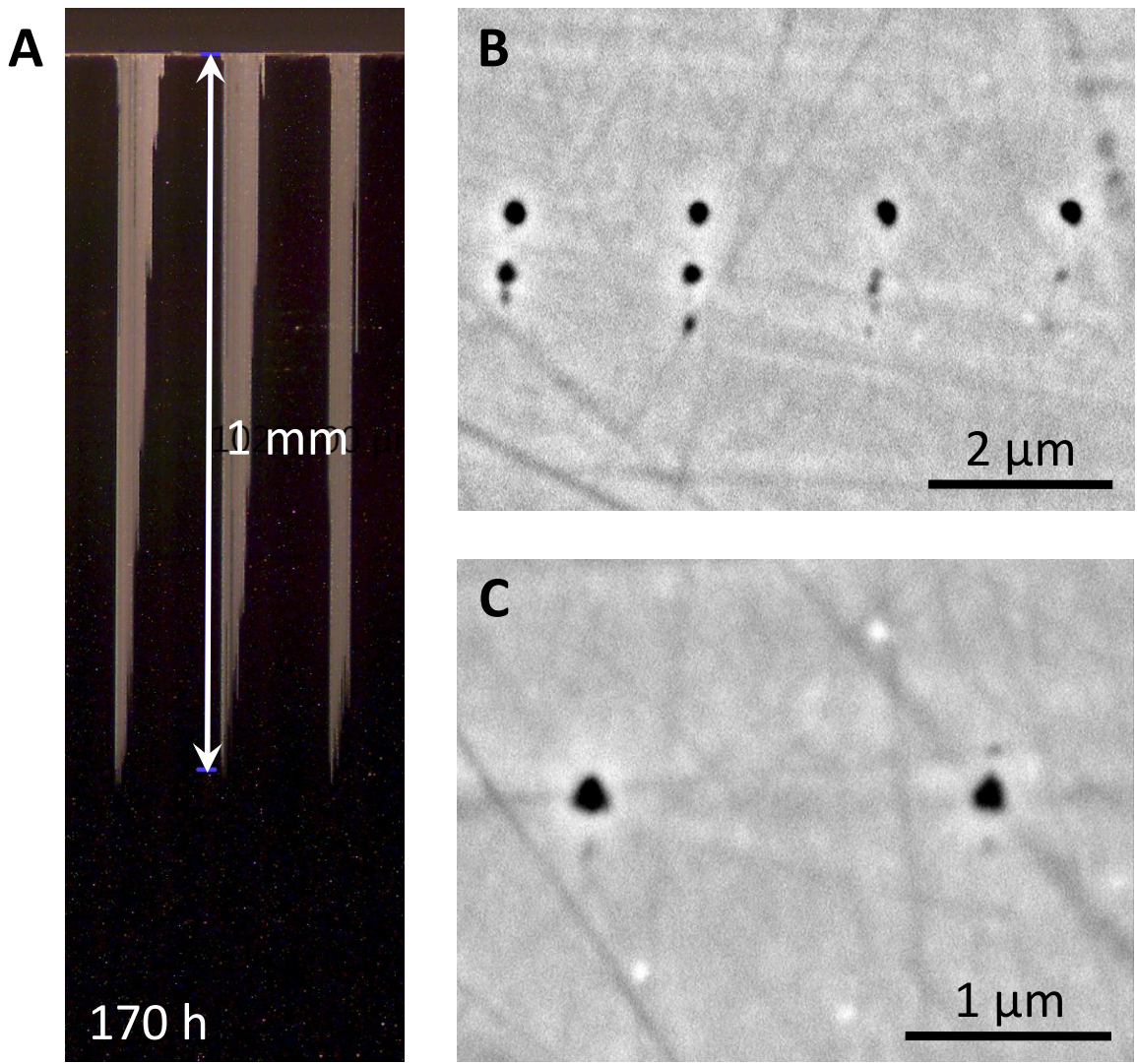

**Fig. S7. Etching of mm-long nanopores in Sapphire.** (**A**) Dark field image of three arrays of 1-mm long pores after 170h of total etching time. Pores on each array were written at ~10 mW and at depths ranging from 4 to 30 µm. (**B**) Example of pores written at medium power (9.4 mW) and 29 µm depth, after 30 min etching. Birefringence-induced focal splitting results in multiple pores along the (top to bottom) in-depth focusing direction. The four main (top) pores have an average width of 197 nm ±12 nm and height of 227 nm ±9 nm. The two observed second pores have an average width of 154 nm ±5 nm and a height of 164 nm ±7 nm. (**C**) Example of two pores written at 24 µm depth and at the photo-modification power threshold (~4 mW) for which no secondary pores are observed. The pores have an average width of 121 nm ±12 nm and height of 135 nm ±21 nm.

## References

1. S. John, Strong localization of photons in certain disordered dielectric superlattices. *Phys. Rev. Lett.* **58**, 2486 (1987). doi: 10.1103/PhysRevLett.58.2486

2. E. Yablonovitch, Inhibited Spontaneous Emission in Solid-State Physics and Electronics. *Phys. Rev. Lett*. **58**, 2059 (1987). doi:10.1103/PhysRevLett.58.2059

3. R. A. Shelby, D. R. Smith, S. Schultz, Experimental verification of a negative index of refraction. *Science* **292**, 5514, 77-79 (2001). doi:10.1126/science.1058847

4. D. R. Smith, J. B. Pendry, M. C. K. Wiltshire, Metamaterials and negative refractive index. *Science* **305**, 5685, 788-792 (2004). doi: 10.1126/science.1096796

5. K. Jain, C. G. Willson, B.J. Lin, Ultrafast deep UV lithography with excimer lasers. *IEEE Electron Device Lett*. **3**, 3, 53-55 (1982). doi:10.1109/EDL.1982.25476

6. K. McGarvey-Lechable, T. Hamidfar, D. Patel, L. Xu, D. V. Plant, P. Bianucci, Slow light in mass-produced dispersion-engineered photonic crystal ring resonators. *Opt. Express* **25**, 3916-3926 (2017). doi:10.1364/OE.25.003916

7. S. Maruo, O. Nakamura, S. Kawata, Three-dimensional microfabrication with two-photon-absorbed photopolymerization. *Opt. Lett.* **22**, 132-134 (1997). doi:10.1364/OL.22.000132

8. M. Straub, M. Gu, Near-infrared photonic crystals with higher-order bandgaps generated by two-photon photopolymerization. *Opt. Lett.* **27**, 1824-1826 (2002). doi:10.1364/OL.27.001824

9. M. Deubel, G. von Freymann, M. Wegener, S. Pereira, K. Busch, C. M. Soukoulis, Direct laser writing of three-dimensional photonic-crystal templates for telecommunications. *Nat. Mater.* **3**, 444-447 (2004). doi:10.1038/nmat1155

10. S. Wong, M. Deubel, F. Pérez-Willard, S. John, G. A. Ozin, M. Wegener, G. von Freymann, Direct Laser Writing of Three- Dimensional Photonic Crystals with a Complete Photonic Bandgap in Chalcogenide Glasses. *Adv. Mater.* **18**, 3, 265-269 (2006). doi: 10.1002/adma.200501973

11. M. Hermatschweiler, A. Ledermann, G. A. Ozin, M. Wegener, G. von Freymann, Fabrication of silicon inverse woodpile photonic crystals. *Adv. Funct. Mater.* **17**, 14, 2273-2277 (2007). doi:10.1002/adfm.200601074

12. J. K. Gansel, M. Thiel, M. S. Rill, M. Decker, K. Bade, V. Saile, G. von Freymann, S. Linden, M. Wegener. Gold Helix Photonic Metamaterial as Broadband Circular Polarizer. *Science* **325**, 1513 (2009). doi:10.1126/science.1177031

13. J. C. Knight, T. A. Birks, P. St. J. Russell, D. M. Atkin, All-silica single-mode optical fiber with photonic crystal cladding. *Opt. Lett.* **21**, 1547-1549 (1996). doi:10.1364/OL.21.001547

14. P. Russell, Photonic Crystal Fibers. *Science* **299**, 5605, 358-363 (2003). doi:10.1126/science.1079280

15. J. M. Dudley, G. Genty, S. Coen. Supercontinuum generation in photonic crystal fiber. *Rev. Mod. Phys.* **78**, 1135–1184 (2006). doi:10.1103/RevModPhys.78.1135

16. J. G. Rarity, J. Fulconis, J. Duligall, W. J. Wadsworth, P. St. J. Russell, Photonic crystal fiber source of correlated photon pairs. *Opt. Express* **13**, 534-544 (2005). doi:10.1364/OPEX.13.000534
15

17. R. J. A. Francis-Jones, R. A. Hoggarth, P. J. Mosley, All-fiber multiplexed source of high-purity single photons. *Optica* **3**, 1270-1273 (2016). doi:10.1364/OPTICA.3.001270

18. E. N. Glezer, E. Mazur, Ultrafast-laser driven micro-explosions in transparent materials. *Appl. Phys. Lett.* **71**, 882 (1997). doi:10.1063/1.119677

19. S. Juodkazis, K. Nishimura, H. Misawa, T. Ebisui, R. Waki, S. Matsuo, T. Okada, Control over the crystalline state of sapphire. *Adv. Mater.* 18, 11, 1361-1364 (2006). doi:10.1002/adma.200501837

20. Materials and methods are available in the Supplementary Information file.

21. J. Fischer, J. B. Mueller, J. Kaschke, T. J. A. Wolf, A.-N. Unterreiner, M. Wegener, Three-dimensional multi-photon direct laser writing with variable repetition rate. *Opt. Express* **21**, 26244-26260 (2013). doi:10.1364/OE.21.026244

22. M. D. Henry, S. Walavalkar, A. Homyk, A. Scherer, Alumina etch masks for fabrication of high-aspect-ratio silicon micropillars and nanopillars. *Nanotechnology* **20**, 25 (2009). doi:10.1088/0957-4484/20/25/255305

23. J. D. Joannopoulos, P. R. Villeneuve, S. Fan, Photonic crystals: putting a new twist on light. *Nature*, **386**, 143-149 (1997). doi:10.1038/386143a0

24. T. Clausnitzer, T. Kämpfe, E.-B. Kley, A. Tünnermann, A. V. Tishchenko, O. Parriaux, Highly-dispersive dielectric transmission gratings with 100% diffraction efficiency. *Opt. Express* **16**, 5577-5584 (2008). doi:10.1364/OE.16.005577

25. L. R. Meza, S. Das, J. R. Greer, Strong, lightweight, and recoverable three-dimensional ceramic nanolattices. *Science* **345**, 6202, 1322-1326 (2014). doi:10.1126/science.1255908

26. J. Bauer, A. Schroer, R. Schwaiger, O. Kraft, Approaching theoretical strength in glassy carbon nanolattices. *Nat. Mater.* **15**, 438-443 (2016). doi:10.1038/nmat4561

27. A. Banerjee, D. Bernoulli, H. Zhang, M.-F. Yuen, J. Liu, J. Dong, F. Ding, J. Lu, M. Dao, W. Zhang, Y. Lu, S. Suresh. Ultralarge elastic deformation of nanoscale diamond. *Science* **360**, 6386, 300-302 (2018). doi:10.1126/science.aar4165

28. J. Schindelin, I. Arganda-Carreras, E. Frise, V. Kaynig, M. Longair, T. Pietzsch, S. Preibisch, C. Rueden, S. Saalfeld, B. Schmid, J. Y. Tinevez. Fiji: an open-source platform for biological-image analysis. *Nat. Methods* **9**, 676–682 (2012). doi:10.1038/nmeth.2019

29. D. E. Zelmon, D. L. Small, R. Page, Refractive-index measurements of undoped yttrium aluminum garnet from 0.4 to 5.0 μm. *Appl. Opt.* **37**, 4933-4935 (1998). doi:10.1364/AO.37.004933

30. Y. Shimotsuma, P. G. Kazansky, J. Qiu, K. Hirao, *Phys. Rev. Lett.* **91**, 247405 (2003). doi:10.1103/PhysRevLett.91.247405

31. A. Ródenas, G. Zhou, D. Jaque, M. Gu, Direct laser writing of three-dimensional photonic structures in Nd:yttrium aluminum garnet laser ceramics. *Appl. Phys. Lett.* **93**, 151104 (2008). doi:10.1063/1.2998258

32. A. G. Okhrimchuk, V. K. Mezentsev, H. Schmitz, M. Dubov, I. Bennion, Cascaded nonlinear absorption of femtosecond laser pulses in dielectrics. *Laser physics* **19**, 7, 1415-1422 (2009). doi:10.1134/S1054660X09070081

33. A. Ródenas, A. Benayas, J. R. Macdonald, J. Zhang, D. Y. Tang, D. Jaque, A. K. Kar, Direct laser writing of near-IR step-index buried channel waveguides in rare earth doped YAG. *Opt. Lett.* **36**, 3395-3397 (2011). doi:10.1364/OL.36.003395

34. D. Choudhury, A. Ródenas, L. Paterson, F. Díaz, D. Jaque, A. K. Kar, Three-dimensional microstructuring of yttrium aluminum garnet crystals for laser active optofluidic applications. *Appl. Phys. Lett.* **103**, 041101 (2013). doi:10.1063/1.4816338

35. J. Basterfield, The chemical polishing of yttrium iron garnet. *J. Phys. D: Appl. Phys.* **2**, 8, 1159-1161 (1969). doi:10.1088/0022-3727/2/8/414

36. G. Zhou, A. Jesacher, M. Booth, T. Wilson, A. Ródenas, D. Jaque, M. Gu, Axial birefringence induced focus splitting in lithium niobate, *Opt. Express* **17**, 17970-17975 (2009). doi:10.1364/OE.17.017970